\documentclass[preprint]{emulateapj}

\slugcomment{April 19, 2010; Accepted to the Astronomical Journal}

\begin{document}

\title{The Nature of the Strong 24 micron {\sl Spitzer} Source J222557+601148: \\
Not a Young Galactic Supernova Remnant}

\author{Robert A.\ Fesen \& Dan Milisavljevic}
\affil{6127 Wilder Lab, Department of Physics \& Astronomy, Dartmouth
  College, Hanover, NH 03755}

\begin{abstract}

The nebula J222557+601148, tentatively identified by Morris et al.\ (2006) as a
young Galactic supernova remnant (SNR) from {\sl Spitzer} Galactic First Look
Survey images and a follow-up mid-infrared spectrum, is unlikely to be a SNR
remnant based on H$\alpha$, [\ion{O}{3}], [\ion{S}{2}] images and low
dispersion optical spectra.  The object is seen in H$\alpha$ and [\ion{O}{3}]
$\lambda$5007 images as a faint, roughly circular ring nebula with dimensions
matching that seen in 24 $\mu$m {\sl Spitzer} images.  Low-dispersion optical
spectra show it to have narrow H$\alpha$ and [\ion{N}{2}]
$\lambda\lambda$6548,6583 line emissions with no evidence of broad or
high-velocity (v $\geq 300$ km s$^{-1}$) line emissions.  The absence of any
high-velocity optical features, the presence of relatively strong [\ion{N}{2}]
emissions, a lack of detected [\ion{S}{2}] emission which would indicate the
presence of shock-heated gas, plus no coincident X-ray or nonthermal radio
emissions indicate the nebula is unlikely to be a SNR, young or old.  Instead,
it is likely a faint, high-excitation planetary nebula (PN) as its elliptical
morphology would suggest, lying at a distance $\sim 2 - 3$ kpc with unusual but not
extraordinary mid-IR colors and spectrum.  We have identified a $m_{r'} = 
22.4 \pm 0.2$ star as a PN central star candidate.
 
\end{abstract}


\section{Introduction}

Of the 274 Galactic supernova remnants (SNRs) currently known \citep{Green09}, few 
are believed to be relatively young with ages of 3000 years or less.  The list of
young Galactic SNRs includes the historic guest stars of SN 1604 (Kepler's SNR), SN
1572 (Tycho's SNR), SN 1054 (Crab Nebula), SN 1006 (G327.6+14.6), SN 386
(G11.2-0.3), and SN 185 (RCW 86). 

There are also about a dozen remnants which do not have firmly
established ages but are probably less than a few thousand years old.  These
include the recently recognized very young remnant G1.9+0.3 (age $\simeq$ 150
yr; \citealt{Reynolds08}, \citealt{Green08}) and Cassiopeia A (SN $\approx$ 1680;
\citealt{Thor01}, \citealt{Fesen06}).  

With so few young SNRs known, the discovery of even one more young Galactic SNR
is significant.  Young SNRs are of special interest since they offer a host of
details on supernova explosions on finer spatial scales than from extragalactic
SNR investigations, including expansion asymmetries, ejecta abundances, and
clues regarding the nature of the progenitor star and its pre-SN environment.

Hence the discovery of a small symmetric nebula, SSTGFLS J222557+601148
(hereafter J222557), detected in 24 $\mu$m {\sl Spitzer} Galactic First Look
Survey (GFLS) images and tentatively identified as a possible SNR with an age
$\sim 1000$ yr by \citet{Morris06} is worth investigating. Here we present
optical images and spectra which indicate this nebula is likely not a supernova
remnant but instead a faint, high-excitation planetary nebula (PN).

\begin{figure*}
\plotone{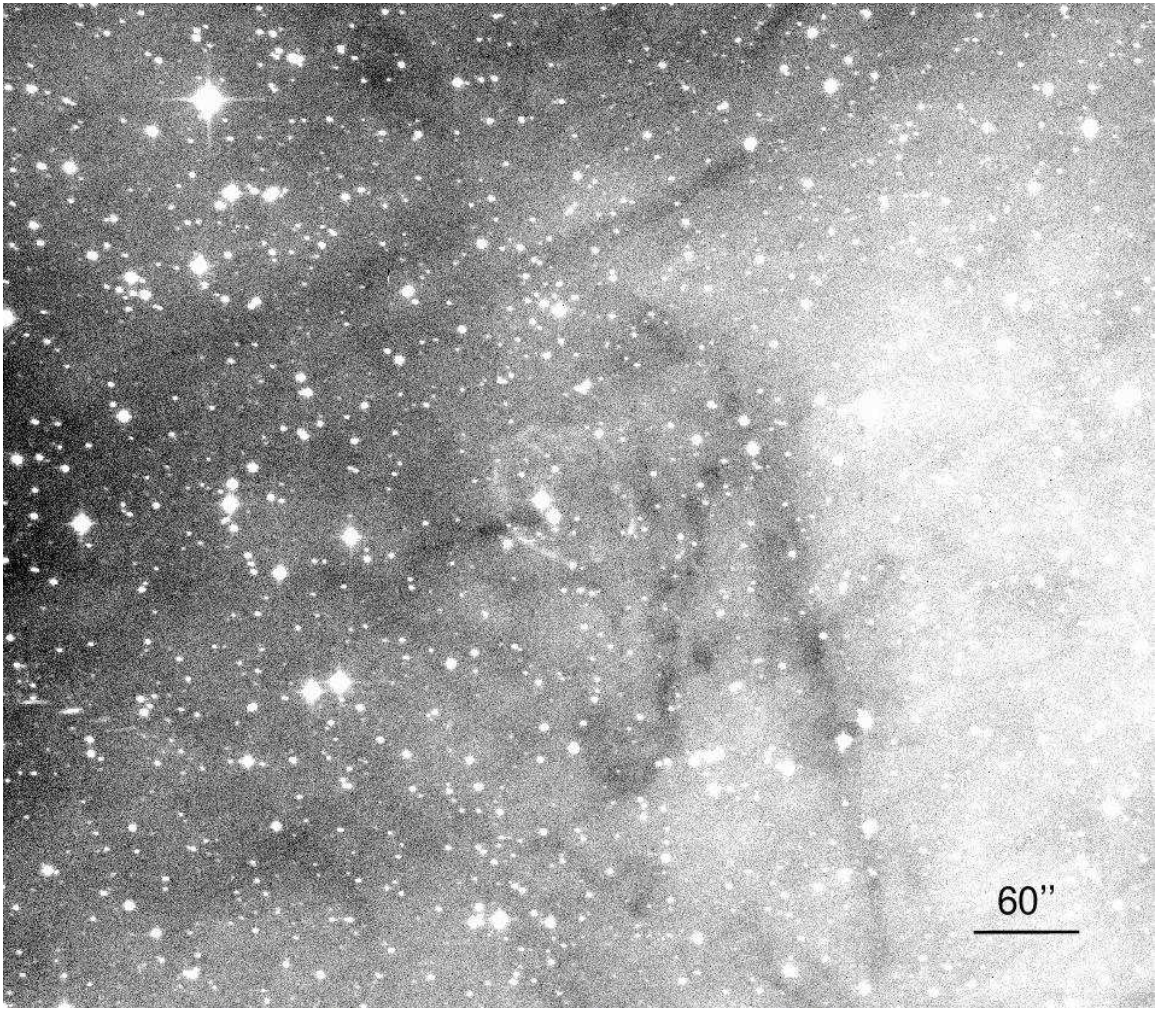}
\caption{H$\alpha$ image of the J222557+601148 region showing the detection of a faint emission shell. 
North is up and East to the left.}
\label{fig:figure0}
\end{figure*}

\begin{figure*}
\plotone{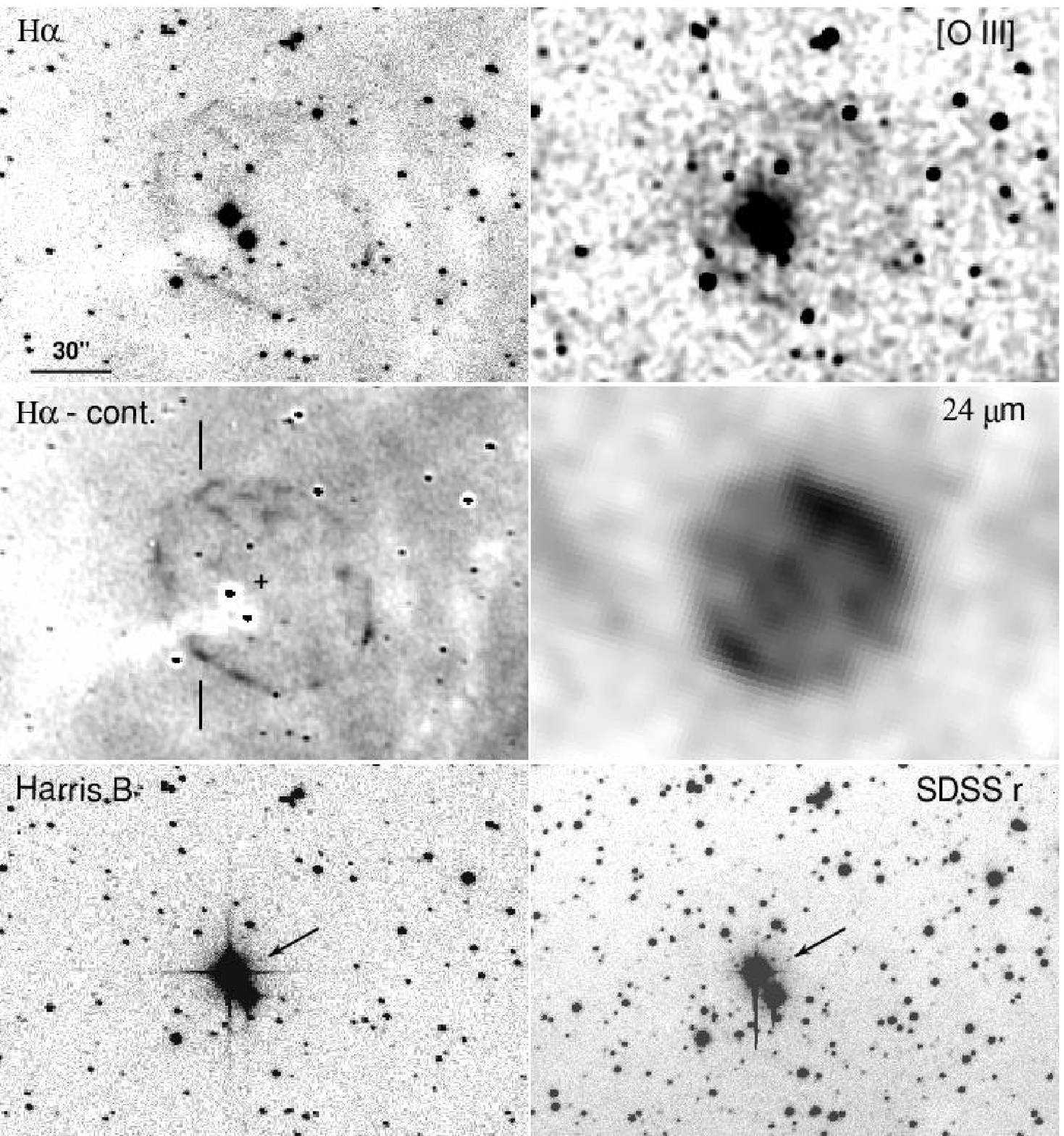}
\caption{Enlarged optical and infrared images of J222557+601148. North is up and East to the left.
{\it{Top row:}} MDM 1.3 m images showing faint H$\alpha$ (left) and [\ion{O}{3}] (right) emission shells. 
{\it{Middle row:}} H$\alpha$ minus 6510 \AA \ continuum (left), and 24 $\mu$m {\sl Spitzer} MIPS
(right). The position where we obtained the optical spectrum using a N--S slit placed 
across the nebula is indicated in the 
H$\alpha$ minus 6510 \AA \ continuum image.
{\it{Bottom row:}} Broadband Harris B and SDSS r filter images with the central star PN (CSPN) candidate 
indicated by the arrows. This star's position relative to the emission shell is shown above in the
H$\alpha$ minus continuum image by a `$+$' symbol. }
\label{fig:images}
\end{figure*}

\section{Observations}

Both narrow and broad passband optical images of J222557 were obtained
in September 2006 using a back-side illuminated 2048 x 2048 SITe CCD detector
attached to the McGraw-Hill 1.3~m telescope at the MDM Observatory (MDM) at Kitt
Peak.  The CCD's 24 micron size pixels gave an image scale of $0\farcs508$ and
a field of view of approximately $17'$ square.  

The nebula was imaged on 2006 Sept 27 using a pair of matched on and off
H$\alpha$ interference filters centered at 6568  
and 6510 \AA \ (FWHM = 30 \AA),  a [\ion{O}{3}] $\lambda$5007 filter
(FWHM = 30 \AA), and a [\ion{S}{2}] $\lambda \lambda$6716,6731 filter (FWHM = 50
\AA).  On Sept 28 broadband images using a Harris B filter were also
taken.  Individual image exposure times for all filters were 1000 s taken in
sets of two or three.

Additional images of J222557 were obtained on 2008 July 30  using the MDM 2.4 m
Hiltner telescope with the RETROCAM CCD camera \citep{Morgan05}.  Images were
taken using SDSS filters $g'$ and $r'$ with exposure times of $2 \times 1200$
and $2 \times 900$ s.  Conditions at the time of the observations were believed
to be photometric with seeing around  1.5$''$.  Images were flux calibrated
with standard stars from \citet{Smith02}. Standard pipeline data reduction of
all images was performed using IRAF/STSDAS\footnote{IRAF is distributed by the
National Optical Astronomy Observatories, which is operated by the Association
of Universities for Research in Astronomy, Inc.\ (AURA) under cooperative
agreement with the National Science Foundation. The Space Telescope Science
Data Analysis System (STSDAS) is distributed by the Space Telescope Science
Institute.}.  This included debiasing, flat-fielding, and cosmic ray and hot
pixel removal.

Follow-up low-dispersion optical spectra of J222557 were obtained on 2008 July
31 using the MDM 2.4 m telescope with a Modular Spectrograph and $2048 \times
2048$ pixel SITe CCD detector.  A north-south 1.2$\arcsec \times 5'$ slit and
a 600 line mm$^{-1}$ 5000 \AA \ blaze grism was used to obtain $2 \times 900$ s
exposures spanning the spectral region $4300 - 7500$ \AA \ with a resolution of
6 \AA.  The slit was placed over the bright southeast region as seen in the 24
$\mu$m {\sl Spitzer} image.  Conditions were good but not photometric with
1.5$''$ seeing.  These spectra were reduced and calibrated employing standard
techniques in IRAF with standard stars from \citet{Strom77}.

\section{Results and Discussion}

As shown in Figure 1, a faint H$\alpha$ emission shell at the location of
J222557 can be seen amidst considerable diffuse H$\alpha$ emission and numerous
dust lanes. In Figure 2, we show enlargements of the H$\alpha$, H$\alpha$ $-$
continuum, and [\ion{O}{3}] $\lambda$5007 images of this optical shell.  The
[\ion{O}{3}] and H$\alpha$ $-$ continuum images are shown smoothed by a 5 point
Gaussian.  The optical shell has angular dimensions of $84'' \times 70''$. This
size is consistent with dimensions of $86'' \times 75''$ reported by
\citet{Morris06} based on the {\sl Spitzer} MIPS 24 $\mu$m image which is also
shown, smoothed by a 3 point Gaussian.  No emission from the shell was seen in
the [\ion{S}{2}] $\lambda\lambda$6716,6731 image despite being of similar or
longer exposure times to those of H$\alpha$ and [\ion{O}{3}].  

\begin{figure}
\plotone{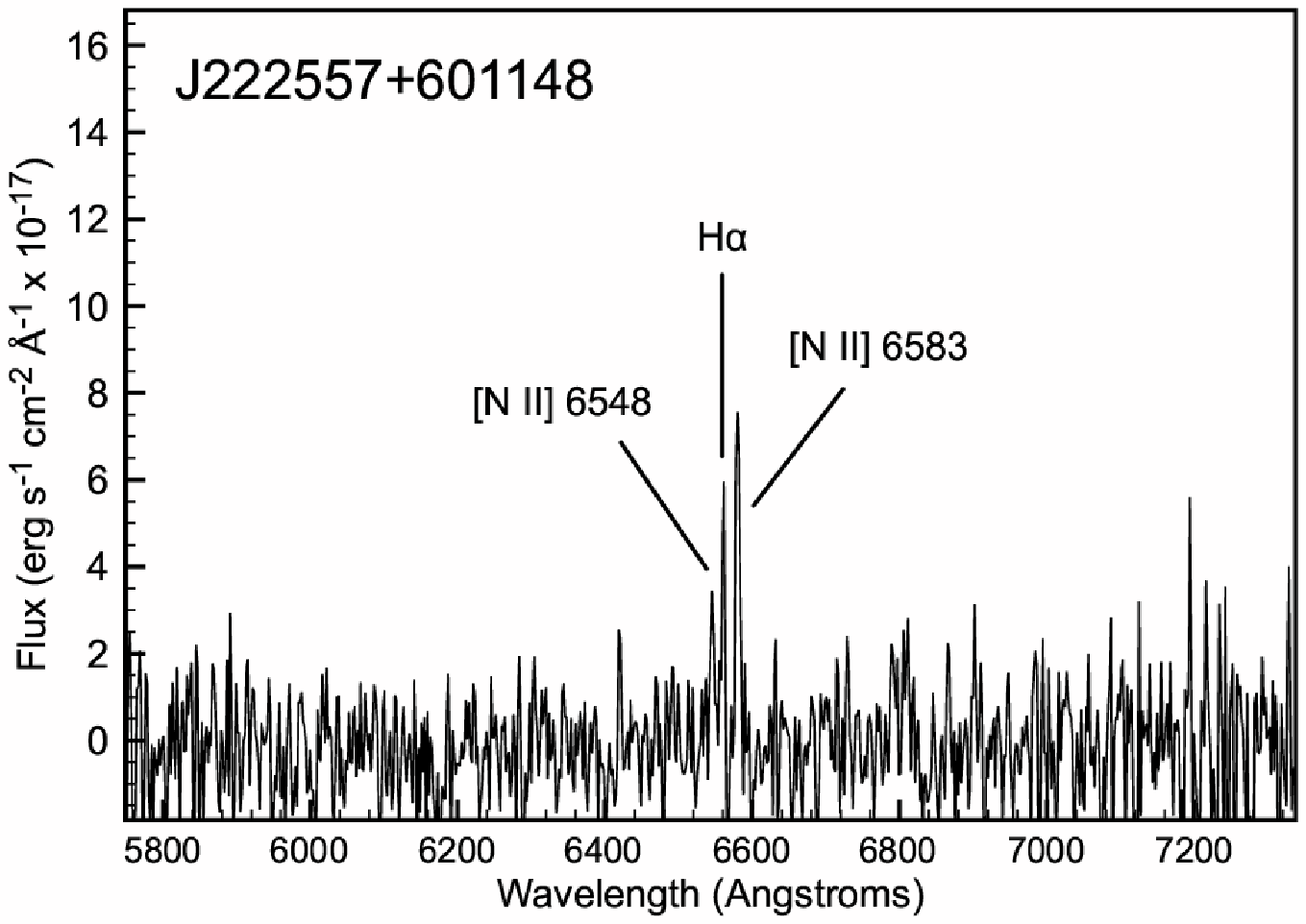}
\caption{Optical spectrum of J222557+601148 showing  
narrow H$\alpha$ and [\ion{N}{2}] $\lambda\lambda$6548,6583 line emissions. }
\label{fig:spectrum}
\end{figure}

In Figure 3, we present our low-dispersion optical spectra of J222557 taken
with a N-S oriented slit positioned along the eastern portion of the ring (see
Fig.\ 2, middle panel).  Narrow, unresolved (FWHM $<$ 6 \AA) emission lines of
H$\alpha$ and [\ion{N}{2}] $\lambda\lambda$6548, 6583 were detected from both
southern and northern limbs.  No high-velocity emission was detected from
either the nebula's shell or from any possible interior emission not visible in
our narrow passband images.  The observed H$\alpha$ flux from the brighter
southern limb knot is approximately $2.4 \times 10^{-16}$ erg s$^{-1}$
cm$^{-2}$ with an uncertainty around 50\% due to non-photometric conditions.
The [\ion{N}{2}] $\lambda$6583 line was stronger than the H$\alpha$ for both
northern and southern portions of the nebula's shell, with I([\ion{N}{2}] 6583
+ 6548)/I(H$\alpha$) $\simeq$ 2.7 for the bright southern knot.

No [\ion{S}{2}] $\lambda\lambda$6716,6731 emission was detected in our spectra
from the nebula, consistent with the lack of any emission seen on the narrow
passband [\ion{S}{2}] image. There was also no [\ion{O}{3}] line emission
detected in the spectrum despite the detected [\ion{O}{3}] $\lambda$5007
emission from the shell in the [\ion{O}{3}] image. We attribute the absence of
detected [\ion{O}{3}] in the spectrum due to a combination of a I([\ion{O}{3}]
$\lambda$5007/I(H$\alpha$) $\lesssim$ 1, considerable line of sight extinction
as suggested by the numerous dust lanes in the region (see Fig.\ 1), and a
lower sensitivity of our spectral setup at $4500 - 5500$ \AA \ compared to the $6000 -
7500$ \AA \ region.  

As noted by \citet{Morris06}, there is no previously known optical, radio, or
X-ray nebula at the location of J222557. The object is not a cataloged
planetary nebula \citep{Acker92} or a known Galactic nebula \citep{NV87}, there
is no coincident radio source in either the Westerbork 325 MHz Northern Sky
Survey (WENSS) or the NRAO VLA Sky Survey 1.4 GHz, and no associated X-ray
emission is visible in the ROSAT All Sky Survey.  The shell is also unlikely to
be a Wolf-Rayet ring nebula due to the lack of a luminous point source in I
band, 2MASS and IRAC images.  Below, we discuss the nature of the J222557
nebula in light of these optical observations, first discussing a young
supernova remnant scenario and then the object as a possible planetary nebula.

\subsection{A Young Supernova Remnant?}

Unusual infrared colors and a pure emission line $7.5 - 37.5$ $\mu$m spectrum
with no sign of dust continuum or hydrogen emission lines led \citet{Morris06}
to examine the possibility of J222557 being a young supernova remnant (SNR).
Although SNRs are usually relatively strong emitters of X-rays and nonthermal
radio emissions, they argued that the object's lack of any such emissions did
not exclude it from being identified as a SNR if the object were relatively
distant and/or if the SNR were relatively young with little interaction with
the ISM.  They noted the nebula's size (dia $\approx 80''$) was consistent
with a young SNR at a distance $\sim 10 $ kpc with an average expansion
velocity of a few 1000 km s$^{-1}$. Such a large distance would also place the
object well above the Galactic plane (z $\simeq$ 400 pc) which would help explain
a lack of ISM interaction and hence its low radio and X-ray luminosity.

As additional support for a SNR identification, \citet{Morris06} cited
J222557's similar morphology and size to the SMC remnant 1E $0102.2-7219$ and
the young Galactic SNR Cassiopeia A (Cas A), both of which are oxygen-rich,
core-collapse remnants.  They noted 1E $0102.2-7219$'s strong 24 $\mu$m
emission and absence in other {\sl Spitzer} IR imaging bands \citep{Stan05}
much like that seen for J222557.  Indeed, many SNRs exhibit strong 24 $\mu$m
emission and a faintness in shorter IR bands and this type of SED is
illustrated in the 3.6, 8.0, and 24 $\mu$m composite color image of the young
galactic core-collapse SNR G11.3-0.3 shown in \citet{Carey09}.

\citet{Morris06} also compared J222557's IRS spectrum with that of Cas A
\citep{Arendt99}. They found similar and consistent line ratios if J222557 had
higher shock velocities than Cas A's $150 - 200$ km s$^{-1}$ in the ejected SN
material of around 450 - 500 km s$^{-1}$.

However, the presence of only narrow H$\alpha$ and [\ion{N}{2}] emissions with
no high-velocity (v $\geq 300$ km s$^{-1}$) emission lines seen in either the
optical or IR, plus no coincident X-ray or nonthermal radio emissions make a
SNR interpretation unlikely. While much of the observed emission structure seen
in our optical images is concentrated in a thin shell where large radial
velocities would not be expected, there is some interior optical emission
(Fig.\ 1) and one of the two {\sl Spitzer} IRS spectra obtained by
\citet{Morris06} was taken of the shell's center and yet showed no hint of
high-velocity.  

Our optical spectrum also revealed no emission from [\ion{O}{1}]
$\lambda\lambda$6300,6364 or [\ion{O}{2}] $\lambda\lambda$7319,7330 which would
be expected if the nebula contained O-rich ejecta as seen in 1E $0102.2-7219$
and parts of Cas A.  Moreover, the absence [\ion{S}{2}]
$\lambda\lambda$6716,6731, which is the dominant line emission in much of Cas
A's ejecta, also greatly weakens an analogy with this remnant.

\citet{Morris06} also discussed the possibility that J222557 might be a
young SNR associated with a thermonuclear Type Ia SN explosion, noting 
that strong 24 $\mu$m emission but weaker emission at $3.6 - 8$ $\mu$m has been
observed in SNRs associated with this type of SN.  For example, \citet{Bor06}
studied four LMC Type Ia remnants and found none were detected in the the {\sl
Spitzer} IRAC bands. \citet{Williams06} found a similar situation in some outer
regions of the LMC remnants N49 and N63A which were attributed to line
emissions from [\ion{O}{6}] 25.9 $\mu$m and [\ion{Fe}{2}] 24.5 and 26.0 $\mu$m.

The morphology of the J222557 nebula certainly resembles the faint thin
shells of young Type Ia SNRs such as Tycho and several young LMC remnants, and
\citet{Morris06} made a cautious comparison of J222557 to the bilaterally
symmetric shell of the Type Ia SN 1006 remnant.  Young Type Ia SNRs exhibit a
spectrum dominated by low-velocity Balmer lines of hydrogen as a result of
high-velocity shocks moving through a partially neutral medium leading to the
production of strong narrow and much weaker broad Balmer lines before complete
ionization occurs \citep{CR78,CKR80,Heng10}.

However, a young Type Ia SNR scenario for J222557 is also inconsistent with its
observed optical spectrum.  Elements other than hydrogen are also collisionally
ionized in the postshock region and may emit line photons, but in the case of
neutral atoms and relatively low-ionization ions, a line's luminosity
is proportional to its ionization time and elemental abundance.  This leads to
relatively weak metal lines compared to the hydrogen Balmer lines.  Thus the 
[\ion{N}{2}] $\lambda$6583 line emissions seen in the J222557 spectrum at
levels actually stronger than H$\alpha$ for both northern and south rims
effectively rules out the optical nebula being due to a fast shock like that
seen in the spectra of young Type Ia SNRs.

Finally, the lack of detected [\ion{S}{2}] $\lambda\lambda6716,6731$ in either our
deep [\ion{S}{2}] image or optical spectra of J222557 is also inconsistent with
the J222557 nebula being an older, more evolved SNR. Due to an extended
postshock cooling zone in remnants with shock velocities below 1000 km
s$^{-1}$, the presence of strong [\ion{S}{2}] emission is one of the chief
identifiers of shock heated gas and has been widely used to discriminate
between photoionized nebulae such as H II regions and planetary nebulae.
Virtually all evolved SNRs show [\ion{S}{2}]:H$\alpha$ $\geq 0.4$ whereas H II
regions typically exhibit values $\leq 0.2$ \citep{Blair81,Fesen85}. 

\subsection{A Planetary Nebula?}

The axial symmetry and limb brightened emission observed in J222557 are
strongly suggestive of a PN nature as \citet{Morris06} themselves discuss.
However, they rejected this identification based on its peculiar infrared
colors, the absence of an IR bright central source often seen in PNe, and its
{\sl Spitzer} IRS spectrum which showed no continuum emission or hydrogen lines
and no dust emission features commonly seen in PNe.  

Nonetheless, a PN nature seems more likely given its optical properties.
Emission line expansion velocities under 100 km s$^{-1}$, relatively weak
[\ion{S}{2}] $\lambda\lambda$6716,6731 emission (I([\ion{S}{2}])/I(H$\alpha) <
0.2$,) and a morphology consistent with a limb brightened shell with some
interior emission knots are properties frequently seen in PNe.  The observed
high I([\ion{N}{2}] 6583 + 6548)/I(H$\alpha$) $= 2.7$ ratio is also often
observed in PNe due to an overabundance of nitrogen in post-MS mass loss
material \citep{Acker89,Acker91}.  

Small ($<1'$) ring and shell morphology nebulae bright at  24 $\mu$m but faint at
3.6 - 8.0 $\mu$m are not unusual in the Galactic plane and some of these are
suspected PNe \citep{Carey09,Flagey09}.  J222557's observed infrared flux of less than 0.1 MJy
sr$^{-1}$ at the four IRAC channels (3.6, 4.5, 5.8, and 8.0 $\mu$m) but about
10 MJy sr$^{-1}$ at 24 $\mu$m indicates a steeply rising SED. This SED is
commonly seen for PNe due in part to significant continuum emission from warm
dust \citep{Zhang09}. The bright appearance of PNe in MIPS 24 $\mu$m images is
due to both dust emission and [\ion{Ne}{5}] 24.3 $\mu$m and [\ion{O}{4}] 25.9
$\mu$m line emissions \citep{Zhang09}. 

The chief difficulty in assigning a PN identification to the J222557 nebula
lies in an absence of appreciable dust continuum emission.  The nebula's
non-detection in the four IRAC 3.6 -- 8.0 $\mu$m images and the 70 or 160
$\mu$m MIPS images \citep{Morris06} sets strong limits on the presence of warm
or cool dust.  The IRS spectrum of J222557 is consistent with the IRAC and MIPS
images, showing a purely emission line spectrum with virtually no continuum or
dust features, but with strong [\ion{O}{4}] 25.9 $\mu$m line emission
explaining its detection in the MIPS 24 $\mu$m image.

However, \citet{Chu09} found that the relative importance of nebular line
emissions and dust continuum emission in the 24 $\mu$m band for PNe depends on
the temperature of the central star and distribution of dust and gas in the
nebula.  Using a sample of 28 PNe, they also found that smaller PNe exhibited
24 $\mu$m emission that was more extended than their H$\alpha$ emission and
concluded that this extended 24 $\mu$m emission was dominated by dust emission.
Larger PNe, in contrast, show much weaker dust emission with the [\ion{O}{4}]
line tending to dominate the emission in the MIPS 24 $\mu$m band.  These
results are in accord with \citet{Stang07} who found the largest PN in the LMC
and SMC exhibited the least dust continuum emission.  

Our optical spectral data are also consistent with J222557 being a PN in terms of its
size.  We measured  v$_{\rm LSR} = -70 \pm 20$ km s$^{-1}$ from the observed
H$\alpha$ line emission, which is consistent with galactic rotation at its $l =
105\fdg8$ and a location inside the Perseus Arm at a distance of $2 - 3$ kpc
\citep{Russeil07}.  With a $b = 2\fdg3$, a distance of $\approx 2.5 $ kpc,
J222557 would lie some $100$ pc above the Galactic plane.  At a distance of 2
to 3 kpc, J222557's angular radius of 40$''$ implies a linear radius of around
$0.5$ pc $\times$ (d/2.5 kpc) which is near the median size of PNe
\citep{Cahn92,Steene94,Ciardullo99,Bensby01,FP10}. Interestingly, some 20\% of
all PNe are not detected in the radio and a size $\geq 0.5$ pc would indicate a
well evolved PN which might help explain J222557's lack of detected radio
emission.
 
Finally, we have identified a possible planetary nebula central star (CSPN)
candidate.  A blueish point source near J222557's center is detected in our
B, $g'$ and $r'$ images, with $m_{g'} = 22.8 \pm 0.2$ and $m_{r'} = 22.4 \pm 0.2$.
Figure 2 (bottom row) shows our Harris B and SDSS $r'$ filter images of the
nebula with the candidate central exciting star indicated.  This star's
location places it nearly centered in the nebula as shown by the cross in the
H$\alpha$ $ - $ continuum difference image.

With an uncertain distance and line of sight extinction, and only very
weak detections in our broadband images, it is difficult to accurately assess
the CSPN candidate's intrinsic color or luminosity. At $m_{V} \approx
22.5$, it would rank among the faintest Galactic CSPN known \citep{Tylenda91}.
However, adopting a distance of 2.5 kpc and an A$_{\rm V} = 4 $ mag based H~I
measurements for $l = 105\fdg8$ and $b = 2\fdg3$ and conversions of N(H) into
A$_{\rm V}$ values for a typical gas to dust ratio \citep{Bohlin78,Pre95}, the
candidate would have M$_{\rm V} \simeq 6.5 $, a value quite consistent for CSPN in
evolved PNe \citep{Phillips05,Benedict09}.  Moreover, given the strength of
[\ion{O}{4}], the nebula is likely to have considerable \ion{He}{2} emission
which would place a lower limit of $\simeq$ 60,000 K for the star's effective
temperature \citep{KJ89}.

\section{Conclusions}

Morris et al. (2006) tentatively identified the Galactic nebula J222557+601148
as a young supernova remnant (SNR) based {\sl Spitzer} Galactic First Look
Survey images. However, H$\alpha$, [\ion{O}{3}], [\ion{S}{2}] images and low
dispersion optical spectra reveal it to only exhibit narrow H$\alpha$ and
[\ion{N}{2}] $\lambda\lambda$6548,6583 line emissions with no evidence of broad
or high-velocity line emissions. 

The absence of any high-velocity optical or infrared features, the presence of
relatively strong [\ion{N}{2}] emissions, a lack of detected [\ion{S}{2}]
emission which would indicate the presence of shock-heated gas, plus no
coincident X-ray or nonthermal radio emissions suggest the nebula is unlikely
to be a SNR, young or old.  Instead, it is likely a faint, high-excitation PN
as its elliptical morphology would suggest, lying at a distance $\sim 2.5$ kpc
with strong [\ion{O}{4}] emission rather than dust continuum emission
dominating it detection in 24 $\mu$m {\sl Spitzer} images.  We have identified
a possible central star candidate with $m_{r'} \simeq 22.4$.

\acknowledgements

We like to thank D. Green for encouraging us to write this work up, and
the anonymous referee for helpful comments which improved the manuscript.

\clearpage
\newpage



\begin{thebibliography}{}
\bibitem[Acker et al.(1989)]{Acker89} Acker, A., Jasniewicz, G., Koeppen, J., \& Stenholm, B.\ 1989, \aaps, 80, 201 
\bibitem[Acker et al.(1991)]{Acker91} Acker, A., Raytchev, B., Koeppen, J., \& Stenholm, B.\ 1991, \aaps, 89, 237 
\bibitem[Acker et al.(1992)]{Acker92} Acker, A., Marcout, J., Ochsenbein, F., Stenholm, B., 
         \& Tylenda, R.\ 1992, Garching: European Southern Observatory, 1992,
\bibitem[Arendt et al.(1999)]{Arendt99} Arendt, R.~G., Dwek, E., \& Moseley, S.~H.\ 1999, \apj, 521, 234 
\bibitem[Benedict et al.(2009)]{Benedict09} Benedict, G.~F., et al.\ 2009, \aj, 138, 1969 
\bibitem[Bensby \& Lundstr{\"o}m(2001)]{Bensby01} Bensby, T., \& Lundstr{\"o}m, I.\ 2001, \aap, 374, 599 
\bibitem[Blair et al.(1981)]{Blair81} Blair, W.~P., Kirshner, R.~P., \& Chevalier, R.~A.\ 1981, \apj, 247, 879
\bibitem[Bohlin et al.(1978)]{Bohlin78} Bohlin, R.~C., Savage, B.~D., \& Drake, J.~F.\ 1978, \apj, 224, 132 
\bibitem[Borkowski et al.(2006)]{Bor06} Borkowski, K.~J., et al.\ 2006, \apjl, 642, L141 
\bibitem[Cahn et al.(1992)]{Cahn92} Cahn, J.~H., Kaler, J.~B., \& Stanghellini, L.\ 1992, \aaps, 94, 399 
\bibitem[Carey et al.(2009)]{Carey09} Carey, S.~J., et al.\ 2009, \pasp, 121, 76 
\bibitem[Chevalier \& Raymond(1978)]{CR78} Chevalier, R.~A., \& Raymond, J.~C.\ 1978, \apjl, 225, L27 
\bibitem[Chevalier et al.(1980)]{CKR80} Chevalier, R.~A., Kirshner, R.~P., \& Raymond, J.~C.\ 1980, \apj, 235, 186 
\bibitem[Chu et al.(2009)]{Chu09} Chu, Y.-H., et al.\ 2009, \aj, 138, 691 
\bibitem[Ciardullo et al.(1999)]{Ciardullo99} Ciardullo, R., Bond, H.~E., Sipior, M.~S., 
         Fullton, L.~K., Zhang, C.-Y., \& Schaefer, K.~G.\ 1999, \aj, 118, 488
\bibitem[Flagey et al.(2009)]{Flagey09} Flagey, N., Billot, N., Carey, S., Noriega-Crespo, A., 
         Shenoy, S., Mizuno, D., Paladini, R., \& Kraemer, K.\ 2009, \baas, 41, 762
\bibitem[Fesen et al.(1985)]{Fesen85} Fesen, R.~A., Blair, W.~P., \& Kirshner, R.~P.\ 1985, \apj, 292, 29 
\bibitem[Fesen et al.(2006)]{Fesen06} Fesen, R.~A., et al.\ 2006, \apj, 645, 283  
\bibitem[Frew \& Parker(2010)]{FP10} Frew, D.~J., \& Parker, Q.~A.\ 2010, Pub.\ Astr. Soc. Australia, in press (arXiv:1002.1525) 
\bibitem[Green et al.(2008)]{Green08} Green, D.~A., 
         Reynolds, S.~P., Borkowski, K.~J., Hwang, U., Harrus, I., \& Petre, R.\ 2008, \mnras, 387, L54 
\bibitem[Green(2009)]{Green09} Green, D.~A.\ 2009, Bulletin of the Astronomical Society of India, 37, 45 
\bibitem[Heng(2010)]{Heng10} Heng, K.\ 2010, Pub.\ Astr. Soc. Australia, 27, 23 
\bibitem[Kalberla et al.(2005)]{Kalberla05} Kalberla, P.~M.~W., Burton, W.~B., 
         Hartmann, D., Arnal, E.~M., Bajaja, E., Morras, R., {\ Pouml}ppel, W.~G.~L.\ 2005, \aap, 440, 775 
\bibitem[Kaler \& Jacoby(1989)]{KJ89} Kaler, J.~B., \& Jacoby, G.~H.\ 1989, \apj, 345, 871 
\bibitem[Morris et al.(2006)]{Morris06} Morris, P.~W., Stolovy, S., Wachter, S., Noriega-Crespo, A., Pannuti, T.~G., 
         \& Hoard, D.~W.\ 2006, \apjl, 640, L179
\bibitem[Morgan et al.(2005)]{Morgan05} Morgan, C.~W., et al.\ 2005, \aj, 129, 2504
\bibitem[Neckel \& Vehrenberg(1987)]{NV87} Neckel, Th., \& Vehrenberg, H.\ 1987 
          ``Atlas of Galactic Nebulae'' (Duesseldorf: Treugesell-Verlag), Vol.\ 2
\bibitem[Phillips(2005)]{Phillips05} Phillips, J.~P.\ 2005, \mnras, 357, 619 
\bibitem[Predehl \& Schmitt(1995)]{Pre95} Predehl, P., \& Schmitt, J.~H.~M.~M.\ 1995, \aap, 293, 889 
\bibitem[Reynolds et al.(2008)]{Reynolds08} Reynolds, S.~P., Borkowski, K.~J., Green, D.~A., Hwang, U., Harrus, I.,
        \& Petre, R.\ 2008, \apjl, 680, L41
\bibitem[Russeil et al.(2007)]{Russeil07} Russeil, D., Adami, C., \& Georgelin, Y.~M.\ 2007, \aap, 470, 161 
\bibitem[Smith et al.(2002)]{Smith02} Smith, J.~A.\ 2002, \aj, 123, 2121
\bibitem[Stanghellini et al.(2007)]{Stang07} Stanghellini, L., Garc{\'{\i}}a-Lario, 
         P., Garc{\'{\i}}a-Hern{\'a}ndez, D.~A., Perea-Calder{\'o}n, J.~V., Davies, 
         J.~E., Manchado, A., Villaver, E., \& Shaw, R.~A.\ 2007, \apj, 671, 1669 
\bibitem[Stanimirovi{\'c} et al.(2005)]{Stan05} Stanimirovi{\'c}, S., Bolatto, A.~D., Sandstrom, K., 
         Leroy, A.~K., Simon, J.~D., Gaensler, B.~M., Shah, R.~Y., \& Jackson, J.~M.\ 2005, \apjl, 632, L103 
\bibitem[Strom(1977)]{Strom77} Strom, K.~M.\ 1977, Kitt Peak National
         Observatory Memorandum, Standard Stars for Intensified Image Dissector
         Scanner Observations (Tucson: KPNO)
\bibitem[Thorstensen et al.(2001)]{Thor01} Thorstensen, J.~R., Fesen, R.~A., 
        \& van den Bergh, S.\ 2001, \aj, 122, 297
\bibitem[Tylenda et al.(1991)]{Tylenda91} Tylenda, R., Acker, A., Raytchev, B., 
         Stenholm, B., \& Gleizes, F.\ 1991, \aaps, 89, 77 
\bibitem[van de Steene \& Zijlstra(1994)]{Steene94} van de Steene, G.~C., \& Zijlstra, A.~A.\ 1994, \aaps, 108, 485 
\bibitem[Williams et al.(2006)]{Williams06} Williams, R.~M., Chu, Y.-H., \& Gruendl, R.\ 2006, \aj, 132, 1877 
\bibitem[Zhang \& Kwok(2009)]{Zhang09} Zhang, Y., \& Kwok, S.\ 2009, \apj, 706, 252 

\end{thebibliography}
\end{document}